\documentclass[12pt,fleqn,a4paper]{article}

\usepackage{a4}
\usepackage{amsfonts}
\usepackage{amssymb}

\newcommand{\beqn}{\begin{eqnarray}}
\newcommand{\eeqn}{\end{eqnarray}}

\newcommand{\vol}{{\cal V}}

\def\Gi{\hat{G}}
\newcommand{\oneone}{e}
\def\nN{\hat{N}}

\newcommand{\Ga}{\alpha}
\newcommand{\Gb}{\beta}

\newcommand{\Ge}{\epsilon}
\newcommand{\Geps}{\varepsilon}

\newcommand{\GTh}{\Theta}

\newcommand{\CL}{{\cal L}}
\newcommand{\CP}{{\cal P}}

\newcommand{\Bphi}{{\overline{\phi}{}}}

\newcommand{\tk}{{\tilde K}}

\newcommand{\ft}[2]{{\textstyle {\frac{#1}{#2}} }}

\newcommand{\dd}{\partial}

\newcommand{\I}{{\rm i}}

\newcommand{\pu}{\partial_{\mu}}
\newcommand{\po}{\partial^{\mu}}
\newcommand{\bi}{{\bar{\imath}}}
\newcommand{\bj}{{\bar{\jmath}}}
\newcommand{\Bi}{{\bar{\imath}}}
\newcommand{\Bj}{{\bar{\jmath}}}

\newcommand{\be}{\begin{equation}}
\newcommand{\ee}{\end{equation}}
\newcommand{\ben}{\begin{displaymath}}
\newcommand{\een}{\end{displaymath}}
\newcommand{\ba}{\begin{eqnarray}}
\newcommand{\ea}{\end{eqnarray}}
\newcommand{\nn}{\nonumber}
\newcommand{\non}{\nonumber\\}
\newcommand{\bean}{\begin{eqnarray*}}
\newcommand{\eean}{\end{eqnarray*}}

\newcommand{\mathon}{\mathversion{bold}}
\newcommand{\mathoff}{\mathversion{normal}}

\def\moth{\mathsurround=0pt}
\newdimen\zo \zo=0pt

\def\tick{\leaders\hrule height 0.5ex depth 0pt \hskip 0.5pt}
\def\upboxfill{$\moth \setbox\zo\hbox{\tick}%
  \hskip 2pt\hbox to 0pt{$\tick$\hss}\hrulefill \hbox to
6pt{$\tick$\hss}$}

\def\dtick{\leaders\hrule height .34pt depth .5ex \hskip 0.5pt}
\def\downboxfill{$\moth \setbox\zo\hbox{\dtick}%
  \hskip 2pt\hbox to 0pt{$\dtick$\hss}\hrulefill \hbox to
6pt{$\dtick$\hss}$}

\newcommand{\la}{\label}

\newcommand{\Ref}[1]{(\ref{#1})}

\makeatletter
\@addtoreset{equation}{section}
\makeatother
\renewcommand{\theequation}{\thesection.\arabic{equation}}

\newcommand{\vl}{{\vphantom{[}}}

\newcommand{\mis}{\!-\!}

\begin{document}

\title{}
\begin{flushright}
\vspace{-3cm}
{\small SPIN-02/46 \\
        ITP-UU-02/70 \\
        ROM2F/2002/33 \\
        hep-th/0212255}
\end{flushright}
\vspace{1cm}

\begin{center}
{\Large\bf Calabi-Yau Fourfolds with Flux \\ [2mm]
 and  Supersymmetry Breaking }
\end{center}

\vspace{0.5cm}

\author{}
\date{}
\thispagestyle{empty}

\begin{center}
{\bf Marcus Berg,}\footnote{e-mail: berg@roma2.infn.it}$^{,\dag}$
{\bf Michael Haack}\footnote{e-mail: haack@roma2.infn.it}$^{,\dag}$
{\bf and Henning Samtleben}\footnote{e-mail: h.samtleben@phys.uu.nl}$^{,*}$
\\
\vspace{0.5cm}
{\it
$^\dag$ Dipartimento di Fisica \\
Universit\'a di Roma, Tor Vergata \\
00133 Rome, Italy
}

\vspace{.3cm}

{\it
$^*$Institute for Theoretical Physics \& Spinoza Institute\\
Utrecht University, Postbus 80.195\\
 3508 TD Utrecht, The Netherlands \\
}

\vspace{1cm}
\end{center}

\begin{center}
{\bf Abstract}
\end{center}

In Calabi-Yau fourfold compactifications of M-theory with flux, we
investigate the possibility of partial supersymmetry breaking in the
three-dimensional effective theory. To this end, we place the effective
theory in the framework of general $N=2$ gauged supergravities, in the
special case where only translational symmetries are gauged. This allows
us to extract supersymmetry-breaking conditions, and interpret them as
conditions on the 4-form flux and Calabi-Yau geometry. For $N=2$ unbroken
supersymmetry in three dimensions we recover previously known results, and
we find a new condition for breaking supersymmetry from $N=2$ to $N=1$,
i.e. from four to two supercharges. An example of a Calabi-Yau
hypersurface in a toric variety that satisfies this condition is provided.

\clearpage

\section{Introduction}

In any attempt to make contact between string/M-theory 
and phenomenology, one has to explain why the world looks
effectively four-dimensional. One possibility would be that the
extra dimensions are compactified on a manifold too small
to be noticed at low energies. Then, in turn, one has to
face the problem that every parameter in the
metric of the internal space shows up as a massless scalar field
in the effective theory --- contradicting experiment.
One solution to this problem that has recently attracted
renewed interest consists in turning on internal background values for
$p$-form field-strengths in the higher-dimensional theory. This leads 
to a gauging of the supergravity in the low energy effective 
action and, quite generically, a potential for the 
scalars of the ``no-scale'' type 
is induced.\footnote{However, stringy $\alpha'$-corrections 
spoil the no-scale structure of the potential \cite{GKP,BBHL}.} 
Thus at least some of the scalar fields become massive. 

From a phenomenological point of view, compactifications with unbroken 
$N = 1$ supersymmetry in four dimensions seem most 
appealing in attempts to approach questions like the hierarchy or 
cosmological constant problem. At the same time, only $N = 1$ 
supersymmetry is consistent with a chiral particle spectrum. 
A scenario leaving this amount of supersymmetry in the context of flux 
compactifications was put forward in \cite{GKP}. 
There, orientifolds of type IIB Calabi-Yau compactifications with 3-form 
flux were studied.
\footnote{For a thorough introduction to orientifolds see
\cite{AS} and the references therein. The scalar potential was also  
investigated in \cite{Michel,TV,Mayr,CKLT,DA,LM}, however in the 
$N=2$ context of Calabi-Yau compactifications without 
orientifolding.} However, due to the presence of O3-planes (and possibly 
D3-branes), the derivation of the effective theory using the Kaluza-Klein 
procedure is not straightforward.

In this note we discuss an alternative, related setting: effective
theories that are related to the type IIB orientifolds with 3-form 
flux, but have a more direct description as Kaluza-Klein 
compactifications. In particular, we 
consider eleven-dimensional supergravity on Calabi-Yau 
fourfolds with 4-form flux.
Such compactifications preserve $N=2$ supersymmetry in three
dimensions,
which is the same amount as $N=1$ in four dimensions.
The precise relation of these fourfold compactifications
to IIB orientifolds can be established as follows.
If the Calabi-Yau fourfold is elliptically fibered, the F-theory 
limit relates the fourfold compactification to type IIB compactified on the 
base of the elliptic fibration \cite{V}. At particular loci of the 
moduli space the type IIB compactifications have a dual description as 
type IIB orientifolds \cite{S}. Furthermore, the 3-form flux of the 
type IIB compactifications stems from the lift of the original 4-form flux 
\cite{GVW,DRS}.

One of the main questions 
in fourfold compactifications that we want to address is whether it is 
possible to break supersymmetry spontaneously from $N=2$
to $N = 1$, i.e.\ from four to two supercharges.
Our strategy to answer this question is the following. 
We will argue that the effective theories from compactifications of 
eleven-dimensional supergravity on Calabi-Yau fourfolds with 4-form flux
are special cases of three-dimensional gauged supergravities, where 
some Peccei-Quinn (PQ) symmetries are gauged (in the case of the 
fourfold compactifications these PQ symmetries stem from the gauge 
symmetries of the 3-form). Using the general results of \cite{dWHS} 
about three-dimensional gauged supergravities it is then
straightforward to find a condition for partial supersymmetry 
breaking.

The effective theories from Calabi-Yau fourfold compactifications 
of eleven-dimensional supergravity with 4-form flux
were previously considered in \cite{BB,PM,GVW,DRS,GSS,HL,BB2,BBHL}.
In \cite{BB}, conditions on the flux for preserving $N=2$ 
supersymmetry were derived. In particular, only a primitive flux of type 
(2,2) preserves $N=2$. It was argued in \cite{GVW} that this
condition can be encoded in two superpotentials, whereas it was shown in 
\cite{HL} that the potential resulting from a Kaluza-Klein 
reduction involves one superpotential and a further 
contribution expressed via a real function closely 
related to the proposed second superpotential of \cite{GVW}.
We show that the correctness of the potential 
found from Kaluza-Klein reduction in \cite{HL}
is confirmed by the general $N=2$ gauged-supergravity potential
given in \cite{dWHS}. More precisely, we
consider the complete bosonic 
part of the action for a three-dimensional supergravity with 
gauged PQ symmetries, in a form that is related to the one derived in 
\cite{dWHS} by a redualization of the PQ scalars to vectors.
The redualized form is well suited for
 the comparison with the theories from fourfold compactifications 
with fluxes.

The potential was used in \cite{BBHL} to show that a 
flux proportional either to the holomorphic 4-form $\Omega$ (and its 
complex conjugate) or the square of the K\"ahler form $J$ 
of the Calabi-Yau classically leads to a vanishing cosmological constant,
despite the complete (spontaneous) breaking of supersymmetry.
This feature had been noticed before in \cite{GVW} (for $F_4 \sim \Omega + c.c.$)
and \cite{BB2} (for $F_4 \sim J \wedge J$). Here we further 
investigate the issue of supersymmetry breaking, and find that 
it is possible to obtain partial breaking from $N=2$ to 
$N=1$ by turning on a certain combination of both of these 
fluxes, i.e. $F_4 = F J \wedge J + (\tilde F \Omega + c.c.)$. 
These three-dimensional $N=1$ vacua might be interesting 
due to Witten's proposal to explain the vanishing of the 
cosmological constant by considering our 
four-dimensional world to be described by the strong coupling limit of an 
$N=1$ supersymmetric three-dimensional theory \cite{W2}.

The article is organized as follows. In section 2 we start with a 
summary of relevant facts about three-dimensional gauged 
supergravity with $N=2$, i.e.\ four supercharges) \cite{dWHS}. This 
theory is formulated with all vectors dualized to scalars,
such that the bosonic spectrum contains only
scalars from chiral multiplets as dynamical degrees of freedom. 
In order to make contact with the effective theory from Calabi-Yau fourfold 
compactifications, we then specialize to the case that some of the  
scalars have a PQ symmetry, which is the 
symmetry that is gauged. For this special 
case we perform a (re-)dualization of the 
PQ scalars to vectors and derive the general form of the bosonic part of 
the effective action. Furthermore, we discuss the condition for unbroken 
$N=2$ and $N=1$ supersymmetry. In section 3 we compare the general form of
the gauged supergravities discussed in section 2 to those from Calabi-Yau 
fourfold compactifications of eleven-dimensional supergravity. 
We start by reviewing the case without 4-form flux, discussed 
before in \cite{HL2}. As opposed to \cite{HL2}
however, here we express the effective action through 
both chiral and vector multiplets to make contact with the results
in section 2. The reason is that it is in these coordinates that the 
flux-induced potential can be expressed most easily. This is done in 
the following subsection, where we turn on 4-form 
flux. We generalize the analysis of \cite{HL} by including the 
moduli from expanding the 
3-form in a basis of
3-forms on the Calabi-Yau, finding that the potential 
is not modified in this case. Furthermore, we make some observations about the 
range of validity of our derivation. We end section 3 with a discussion of 
partial supersymmetry breaking to $N=1$, specializing the general 
condition from section 2 to the case at hand. By giving a concrete (toy) example 
we show that it is indeed possible to fulfill this condition. Finally, an
appendix contains some more details  
on the search for an example of partial supersymmetry breaking.


\section{The gauged supergravity}

In this section, we first review the general form of $N=2$
gauged supergravity in three dimensions as given in~\cite{dWHS},
generalizing the abelian $U(1)$ gaugings of~\cite{Deger:1999st,AZS}. We then
turn to gauge groups that are exclusively generated by translational
(Peccei-Quinn) symmetries of the scalar target space and show that
these theories admit a dual formulation in which the PQ scalars are
redualized into vector fields. From the supersymmetry transformation
rules, we eventually derive the conditions of $N=1, 2$ preserved
supersymmetries at a stationary point of the scalar potential.

\mathon
\subsection{$d=3$, $N=2$ supergravity}
\mathoff

We start from the formulation of ungauged $d=3$, $N=2$
supergravity in which all vector fields are dualized into scalars.
The total scalar target space is a K\"ahler manifold which we
parametrize by complex coordinates $\{\phi^i,\Bphi{}^\bi\}$ with the
metric derived from the K{\"a}hler potential $g_{i\bi} = \dd_i \dd_\bi
K$\,. An isometry is described by a holomorphic vector field $\dd_\bi
X^j=0$, satisfying
\ba
\nabla_i X_\bj + \nabla_\bj X_i &=& 0 \;.
\ea
Gauging a subgroup of isometries that is generated by a basis $\{ X^{
j}_A \}$ corresponds to defining space-time vector fields
$A_{A\,\mu}$ via the duality equation\footnote{We take the signature 
of the space-time metric to be $(+--)$. The
$\epsilon_{\mu\nu\rho}$ is defined to transform as a
tensor, i.e. $\epsilon_{123}=e$, $\epsilon^{123}=e^{-1}$. 
Our curvature conventions are
$R_{\mu\nu\sigma}{}^{\tau}=\dd_\nu\Gamma^\tau_{\mu\sigma} +
\Gamma^\tau_{\nu\lambda}\,\Gamma^{\lambda}_{\mu\sigma} -
(\mu\leftrightarrow\nu),\,R_{\mu\nu}=R_{\mu\tau\nu}{}^{\tau}$.}
\ba
\Ge_{\mu\nu\rho}\,F_A^{\nu\rho} &=&
-4 \Re\!\left[g_{i\Bi} X_A^\Bi D_\mu\phi^i\right] + \mbox{fermions}\;,
\la{duality}
\ea
and minimally coupling them to the scalar fields by means of a
constant symmetric matrix $\Theta^{AB}$ as
\ba
D_\mu \phi^j &=& \partial_\mu \phi^i + g \Theta^{AB}\,A_{A \mu}
(X^{i}_B \dd_i+ X^{\Bi}_B \dd_\Bi)\,\phi^j \;.
\la{Dcov}
\ea
The duality equation \Ref{duality} consistently implies the scalar
equation of motion as integrability condition. Supersymmetry further
requires the introduction of fermionic mass terms, a Chern-Simons
term for the vector fields that gives rise to \Ref{duality} as field
equation, and a scalar potential. The gravitino mass tensor is given
by
\ba
A_1 &=& \left(\begin{array}{rr} 2T & 0 \\ 0 & 2T \end{array}
\right) 
+ e^{K/2}\,
  \left(
\begin{array}{rr} -\Re W & \Im W \\ \Im W & \Re W \end{array}
\right) \;,
\la{A1}
\ea
with $T \equiv \CP\!_A \GTh^{AB} \CP\!_B$, where the real function
$\CP\!_A$ is the momentum map associated with the isometry $X_A^i$
\ba
\dd_j \CP\!_A &\equiv& \frac12 \I\,g_{j\bj}X^{\bj}_A
\;.
\la{VX}
\ea
The gauging is uniquely determined by the choice of $\Theta^{AB}$
up to the superpotential $W$ which is an arbitrary holomorphic
function of the scalar fields, subject to
\ba
X^{i}_A  D_i W
&=&
2\,\I\, W {\CP}\!_{A}\;.
\la{constraint}
\ea
with the K{\"a}hler covariant derivative $D_i W \equiv \dd_i W + \dd_i
K\,W $\,. The complete bosonic part of the Lagrangian is given
by
\ba
e^{-1}\CL_g &=& \ft12 R -g_{i{\Bj}}\,D_\mu \phi^i D^\mu
\Bphi{}^{\Bj} -\ft14
g\GTh^{AB}\epsilon_{\mu\nu\rho} A_A^\mu F_{B}^{\nu\rho}
 +  V \;,
\la{Lold}
\ea
with a scalar potential 
\ba
V &=& g^2 \left(
 4g^{i\bi}\dd_iT\dd_\bi T -4T^2 + e^K (
\ft14\,g^{i\bi}D_i W D_\bi \overline{W} - |W|^2 )\right) \;.
\la{potential}
\ea
Summarizing, the gauged deformation of the $N=2$ supergravity is
uniquely determined by the constant matrix $\GTh^{AB}$ describing the
minimal coupling \Ref{Dcov}, and a holomorphic superpotential $W$
satisfying \Ref{constraint}. It further induces fermionic mass terms
as well as a Chern-Simons term and the scalar potential in the
Lagrangian. Note that after gauging, the scalar fields may
in general no longer be redualized into vector fields.

\subsection{Redualizing}

Let us now specialize this construction to gauge groups that are
generated by translational symmetries of the scalar target space. As
it turns out, in these theories the charged scalar fields may be
redualized into vector fields and completely  removed from the
Lagrangian. Their removal turns the formerly nonpropagating vector
fields into physical fields that appear with a Yang-Mills kinetic term
and an additional Chern-Simons coupling in the Lagrangian. These are
the theories that describe the compactifications of
eleven-dimensional supergravity on Calabi-Yau fourfolds with
background fluxes.

We introduce the following split of scalar fields
\ba
\phi^i &=& \left( \phi^a ,\, \phi^A \right) \;,
\la{split}
\ea
and define the real components $\varphi^A=\Re \phi^A$,
$\hat{\varphi}^A=\Im \phi^A$. We now assume that
the K\"ahler potential does not depend on the $\hat{\varphi}^A$
\ba
K &=& K(\phi^a ,\, \varphi^A) \;.
\la{Krest}
\ea
In particular, this implies that
\ba \label{gij}
g_{i\Bj} &\equiv&
\left(
\begin{array}{cc}
g_{a{\bar{b}}} & g_{a B} \\[.5ex]
g_{A \bar{b}} & g_{AB}
\end{array}
\right)
\;,
\ea
with $g_{a{\bar{b}}} = \dd_a \dd_{{\bar{b}}} K$,\, $g_{a A}=
\ft12\dd_a\dd_A K$,\, and the real matrix $g_{AB} =\ft14 \dd_A
\dd_B K \equiv \ft12 G_{AB}$. The partial derivatives $\dd_A$ here
always refer to real derivatives w.r.t.\ the field $\varphi^A$ in
\Ref{Krest}. To avoid confusion, small~$g$ will always refer to the
metric obtained from the original K\"ahler potential $K$ in the
completely dualized picture in which everything is expressed in chiral
multiplets.

The geometry described by \Ref{Krest} is invariant under
shifts in $\hat{\varphi}^A$, i.e.\ the metric~\Ref{gij} admits a set
of ``elementary'' isometries, labelled by a subscript~$B$, which are
generated by the constant imaginary vector fields
\ba
X_B^i &=& (0 , \I \delta_B^A ) \;,
\ea
in the basis \Ref{split}. From \Ref{VX} we find that the associated
momentum map is given by
\ba
\CP_A &=& \ft14\, \dd_A K \;.
\la{mm}
\ea
The most general gauging of these shift
isometries is described by a constant matrix $\GTh^{AB}$ such that the
covariant derivatives become
\ba
D_\mu \hat{\varphi}^A &=& \dd_\mu \hat{\varphi}^A
+ g \GTh^{AB} A_{B\,\mu} \;,
\nn
\ea
where the vector field is defined by the duality equation
\Ref{duality}
\ba \la{fdualize}
F_{A\,\mu\nu} &=&
- \epsilon_{\mu\nu\rho}
\left(G_{AB} \,(\dd^\rho\hat{\varphi}^B +
g \GTh^{BC} A_C^{\rho} ) +
2\,\Im\left[g_{a A}\,\dd^\rho\phi^a\right] \right)
\;.
\ea
Let us assume that the submatrix $G_{AB}$ of $g_{i\Bi}$ is
separately invertible with inverse $G^{AB}$. Then
$\dd^\rho\hat{\varphi}^B$ may be expressed in terms of
$F_{A\,\mu\nu}$, and we derive that
\ba
D^\mu (G^{AB} F_{B\,\mu\nu}) &=&
\ft12g\GTh^{AB}\epsilon_{\nu\mu\rho} F_{B}^{\mu\rho}
+ 2\epsilon_{\nu\mu\rho}
\,\Im\left[
\dd^\mu\!\left(G^{AB} g_{a B}\,\right)
\dd^\rho\phi^a \right] \;.
\nn
\ea
That is,
the scalar fields $\hat{\varphi}^A$ can be eliminated from all equations of
motion. In turn, the vector fields now satisfy a second order field
equation which may be derived from the Lagrangian
\ba
\ft14\, G^{AB} F_{A}^{\mu\nu} F^\vl_{B\,\mu\nu}
+\ft14\,
g\GTh^{AB}\epsilon_{\mu\nu\rho} A_A^\mu F_{B}^{\nu\rho}
+ \epsilon_{\mu\nu\rho}  F_A^{\mu\nu}\,
G^{AB}\, \Im\left[ g_{a B}
\dd^\rho\phi^a \right] \;.
\nn
\ea
The remaining kinetic terms of the Lagrangian after redualization are
given by
\ba
-\ft12\,G_{AB}\dd_\mu\varphi^A\dd^\mu\varphi^B
- 2\Re\left[ g_{a A}\dd_\mu\varphi^A\dd^\mu\phi^a \right]
- \left(g_{a\bar{b}} -
g_{a A} G^{AB} g_{\bar{b} B}   \right)
\dd_\mu\phi^a \dd^{\mu}\Bphi{}^{\bar{b}}
\non
{}-\ft12\,g_{a A} G^{AB} g_{b B}  \dd_\mu\phi^a
\dd^{\mu}\phi^b -
\ft12\,g_{\bar{b} A} G^{AB} g_{\bar{b} B}
\dd_\mu\Bphi{}^{\bar{b}} \dd^{\mu}\Bphi{}^{\bar b}
\;.
\nn
\ea
Defining new real coordinates
\ba \label{ma}
M_A &\equiv& \ft12\,\dd_A K \;.
\ea
such that $\dd_A M_B = G_{AB}$,\,\,$\dd_a M_A = g_{a
A}$, the complete Lagrangian after redualization then takes the form
\ba
e^{-1}\CL_g &=& \ft12 R -\ft12\,G^{AB}\,\dd_\mu M_A \dd^\mu M_B
+ \ft1{4}\, G^{AB} F_{A}^{\mu\nu} F^\vl_{B\,\mu\nu}
-G_{a{\bar{b}}}\,\dd_\mu \phi^a \dd^\mu
\Bphi{}^{\bar{b}}
\non[1ex]
&&{}
+ \epsilon_{\mu\nu\rho}  F_A^{\mu\nu}\,
\Im\left[ G^{AB} g_{a B} \,
\dd^\rho\phi^a \right]+\ft14
g\GTh^{AB}\epsilon_{\mu\nu\rho} A_A^\mu F_{B}^{\nu\rho}
 +  V \ ,
\la{Lnew}
\ea
with the scalar potential $V$ from \Ref{potential} above and the
metric
\ba
G_{a{\bar{b}}} &\equiv&
(G^{a{\bar{b}}})^{-1} ~\equiv~
(g^{a{\bar{b}}})^{-1} ~=~
g_{a{\bar{b}}} - 2 g_{a A} G^{AB} g_{\bar{b} B} \;.
\nn\label{Gg}
\ea
To clarify the last equation, we note that the inverse of $g_{i\Bj}$
(\ref{gij}) is given by
\beqn \label{ginv}
g^{i\Bj}
&=&
\left(
\begin{array}{cc}
g^{a\bar{b}} &
- 2g^{a\bar{b}} g_{\bar{b} B} G^{BA} \\[.5ex]
- 2g^{a\bar{b}} g_{\bar{b} B} G^{BA} &
2G^{AB} + 4 G^{AC} g_{a C} g^{a\bar{b}}
g_{\bar{b} D} G^{DB}
\end{array}
\right),
\eeqn
where $g^{a\bar{b}}=\left(g_{a{\bar{b}}} - 2 g_{a A} G^{AB} g_{\bar{b}
B}\right)^{-1}$. Hence, the metrics which appear in the kinetic terms
of the dualized Lagrangian~\Ref{Lnew} are not part of the blocks of
the original K\"ahler metric~\Ref{gij} or its inverse \Ref{ginv}, but
are instead given by
\ba
G_{AB}=2g_{AB}\;, \quad G^{AB}\not=\ft12g^{AB}\;, \quad
G_{a{\bar{b}}}\not=g_{a{\bar{b}}}\;, \quad
G^{a{\bar{b}}}=g^{a{\bar{b}}}
\;.
\nn
\ea
We should emphasize that these metrics as well as the mixed coupling
in the dualized Lagrangian (\ref{Lnew}) can all be expressed via a
``kinetic potential'' $\tk(\phi^\alpha,M_A)$ as
\be \label{dtildek}
G^{AB}  =  - \ft12\, \dd_{M_A} \partial_{M_B} \tk \;,\quad
G_{a\bar{b}} = \dd_{\phi^a}
\partial_{\phi^{\bar{b}}} \tk\;,\quad
G^{AB} g_{a B} =  \ft12\,
\dd_{\phi^a} \partial_{M_A} \tk\;,
\ee
where $\tk$ is given by
\be \label{ktilde}
\tk = K - 2 M_A \varphi^A\ \;.
\ee
Moreover, the original K\"ahler potential $K$ can be recovered
from $\tk$ as
\be \label{ktk}
K = \tk - M_ A \partial_{M_A} \tk\ .
\ee
The scalar potential $V$ in \Ref{Lnew} finally is expressed in terms
of $T$ and a holomorphic function $W$. From \Ref{mm} we find that the
consistency constraint \Ref{constraint} is simply equivalent to
assuming that $W=W(\phi^\alpha)$ is a holomorphic function of the
fields $\phi^\alpha$ only. In the new coordinates $(\phi^a, M_A)$, the
potential~\Ref{potential} simplifies to
\ba
g^{-2} V &=&
\ft12 M_A  \GTh^{AC} G_{CD} \GTh^{DB} M_B
-\ft14\left(M_A\GTh^{AB}M_B\right)^2
\non[.5ex]
&&{}
+\ft14\,e^K\,G^{a\bar{b}}
D_a W D_{\bar{b}} \overline{W}
- e^K \left(1- \ft12 M_A G^{AB} M_B  \right) |W|^2
\la{Vnew}\;,
\ea
with the original K\"ahler potential $K$, a holomorphic function
$W(\phi^\alpha)$ and the K\"ahler covariant derivatives $D_a W=\dd_a W
+(\dd_a K) W$.  Note that the metrics $G^{a \bar b}$, $G_{AB}$ 
appearing here are precisely the inverses of the metrics
in the kinetic terms in \Ref{Lnew}, and that all mixed terms vanish. 
Furthermore, the
potential can be expressed entirely in terms of the kinetic
potential (\ref{ktilde}), using (\ref{dtildek}) and (\ref{ktk}).  Let
us finally note that for $M_A G^{AB} M_B=2$, the last term in
\Ref{Vnew} vanishes, and the potential is in fact positive definite
\ba
g^{-2} V &=&
\ft12\left( M_A \GTh^{AC} \mis 2T M_A G^{AC} \right)
G_{CD} \left(\GTh^{DB}M_B \mis 2T G^{DB} M_B \right)
\non
&&{}
+\ft14e^K G^{a\bar{b}}
D_a W D_{\bar{b}} \overline{W}
\;.
\la{posdef}
\ea

Summarizing, we have seen that in the special case of a gauging of the
Peccei-Quinn symmetries associated with some of the scalar fields,
these fields may be dualized into vector fields. In the dual picture
the physical fields are the complex scalars $\phi^a$, the real scalars
$M_A$, and the vector fields $A_{A\,\mu}$; their dynamics is
described by the Lagrangian \Ref{Lnew} with scalar potential
\Ref{Vnew}. This may be viewed as a deformation of the Lagrangian
$\CL_{g=0}$ that is triggered by the Chern-Simons term for the vector
fields (which encodes the constant matrix $\GTh^{AB}$) whose
supersymmetrization then induces the fermionic mass terms and the
scalar potential. Again, this deformation is unique up to the choice
of the holomorphic superpotential $W$.

A noteworthy property of the dualized Lagrangian is the appearance of
the kinetic potential $\tk$ in the kinetic terms of (\ref{Lnew}),
whereas the part of the potential \Ref{Vnew} derived from a
superpotential still involves the old K\"ahler potential $K$, both via
the overall factor $e^K$ and the K\"ahler covariant derivatives. Let
us remind the reader that a similar situation occurs in $D=4$ in an
ungauged theory with both chiral and linear multiplets. (The latter
become vector multiplets in a reduction to three dimensions and
therefore the similarity is no accident.)  Namely, consider the $D=4$
theory described by a K\"ahler potential~\cite{BGG,ABGG}
\be \label{kahler4d}
K(\phi^a, \bar \phi^{\bar a}, L) =
K_0 (\phi^a, \bar \phi^{\bar a}) + \alpha \ln L\ ,
\ee
where $\phi^a$ denote the (complex) scalars from the chiral
multiplets, $L$ is one (real) scalar from the linear multiplet, and
$\alpha$ is an arbitrary constant.\footnote{We use the term K\"ahler
potential here although, like in the three-dimensional case, the
scalar manifold is only K\"ahler when expressed in the dual, chiral
variables.}  The motivation to look at this special form of the
K\"ahler potential in $D=4$ comes from the fact that it is exactly
what one needs to describe the linear dilaton multiplet in heterotic
string models. The effective action of the linear multiplet coupled to
chiral multiplets involves another real function of the $\phi^a$, the
'linear potential' $V(\phi^a, \bar \phi^{\bar a})$. In the context of
the heterotic string this encodes the (non-holomorphic) threshold
corrections to the gauge coupling and in the present
context it is the analog of $\varphi$ in (\ref{ktilde}). More
precisely, defining the kinetic potential
\be \label{kti}
\tk = K - 3 L V \;,
\ee
the bosonic part of the Lagrangian including a potential derived from
a superpotential takes the form \cite{BGG}
\beqn \label{4dact}
e^{-1} {\cal L} & = &  \frac12 R - \tk_{a \bar a}
\partial_\mu \phi^a \partial^\mu \bar \phi^{\bar a} +
\frac14 \tk_{LL} \partial_\mu L \partial^\mu L  \non
& &
\mbox{} - \frac{1}{3!} \tk_{LL}
H_{\mu \nu \rho} H^{\mu \nu \rho} + \frac{1}{3!}
\epsilon_{\mu \nu \rho \sigma} H^{\nu \rho \sigma}
\Im (\tk_{La} \partial^\mu \phi^a)  \non[.5ex]
& & \mbox{}
+ e^K \Big[ \tk^{a \bar a} D_a W D_{\bar a} \bar W
- (3 - L K_L) |W|^2 \Big] \;,
\eeqn
where the covariant derivatives $D_a W = (\partial_a + K_a) W$ involve
the old K\"ahler potential.\footnote{We have adopted the signs in the
formula of \cite{BGG} to our choice of conventions.} Taking into
account that (\ref{ktk}) implies
\ben
M_A M_B (\partial_{M_A} \partial_{M_B} \tk) = -M_A (\partial_{M_A} K)\ ,
\een
the similarity between (\ref{4dact}) and (\ref{Lnew}),
(\ref{Vnew}) is obvious, up to the terms depending on
$\Theta^{AB}$ which correspond to the gauging in three dimensions.

\subsection{Supersymmetry breaking}

The fermionic part of the Lagrangian~\Ref{Lnew} and supersymmetry
transformation rules may be extracted from~\cite{dWHS} upon eliminating
$\hat{\varphi}$ by means of \Ref{fdualize}. Here, we restrict to giving
the supersymmetry variation of the gravitino, which is sufficient to
determine the conditions of $N=1, 2$ preserved supersymmetries
at a stationary point of the scalar potential. From~\cite{dWHS}, we
infer that 
\ba
\delta\psi^I_\mu &=& \nabla\!_\mu \Ge^I
+ \I g A_1^{IJ} \gamma_\mu \,\Ge^J
+ \Big( \ft12 \Re \left[ \dd_\mu \phi^i \dd_i K \right]
+ g\GTh^{AB} A_{A\mu} \CP\!_B  \Big) \, \Geps^{IJ} \,\Ge^J
\;,
\la{susy}
\ea
where the index $I=1, 2$ labels the supercharges and $\Geps^{IJ}$ is
the antisymmetric tensor $\Geps^{12}=1$. The Killing spinor associated
with a supersymmetric ground state of the potential
\Ref{Vnew} (at which we may set $\dd_\mu \phi^i=0=A_{A\mu}$) is then
given by tensoring a three-dimensional AdS/Minkowski Killing spinor
with an eigenvector of $A_1$. Vanishing of \Ref{susy} imposes the
relation $g |\lambda| = \sqrt{-V_0}$ between the corresponding
eigenvalue $\lambda$ of $A_1^{IJ}$ and the value of the potential
$V_0$ at the critical point. With the eigenvalues of
\Ref{A1} given by
\ba
\lambda_{\pm} &=& -\ft12\,M_A\GTh^{AB}M_B \pm e^{K/2} |W|  \;,
\nn
\ea
this condition translates into
\ba
\pm 2 e^{K/2} M_A\GTh^{AB}M_B\, |W|  &\equiv &
M_A  \GTh^{AC} G_{CD} \GTh^{DB} M_B
+ e^K M_A G^{AB} M_B \, |W|^2
\non[1ex]
&&{}
+\ft12\,e^K\,G^{a\bar{b}}
D_a W D_{\bar{b}} \overline{W} \;.
\la{N1}
\ea
In order to preserve the full $N=2$ supersymmetry, i.e.\
satisfy this equation for both $\lambda_+$ and $\lambda_-$, both sides
of this equation must vanish, thus implying the conditions
\ba \label{n2}
\GTh^{AB}M_B &\equiv&0\;,\quad W ~\equiv~0 \;,\quad
D_a W ~\equiv~ 0  \;.
\ea
This obviously also implies stationarity of the potential
\Ref{Vnew} as well as vanishing of the cosmological constant. For an
extremum with $N=1$ supersymmetry, it remains to reconcile
stationarity of \Ref{Vnew} with the condition \Ref{N1} for one choice
of sign. As a special case, we may consider the positive definite
potential~\Ref{posdef}, for which $V_0=0$ is a sufficient condition
for a minimum. The latter is equivalent to
\ba\la{n1a}
\GTh^{DB}M_B \mis 2T G^{DB} M_B &=& 0 \;,\qquad D_a W ~=~ 0 \;,
\ea
while the condition~\Ref{N1} in this case reduces to
\ba \label{n1}
M_A\GTh^{AB}M_B &=& \pm 2 e^{K/2}\, |W| \;.
\ea
The two conditions~\Ref{n1a}, \Ref{n1} are necessary and sufficient
for the existence of a minimum with $N=1$ supersymmetry and
vanishing cosmological constant in the positive definite
potential~\Ref{posdef}.  


\section{Kaluza-Klein reduction}

In this section we want to make contact between the general discussion
of the previous section and Calabi-Yau fourfold compactification
of eleven-dimensional supergravity.
We start by reviewing the case without flux and then 
consider the modifications in its presence. Without flux
the reduction leads to an ungauged supergravity, whose 
scalar manifold has some PQ symmetries. These are then  
gauged by turning on the flux.

\subsection{Compactification without flux}

The compactification without flux was investigated
in \cite{HL2}.\footnote{Here we extend however the results of
\cite{HL2} in that we make contact to the formulas derived in the
last section, i.e.\ we formulate the theory using both, chiral and
vector multiplets.} The starting point is the eleven-dimensional
supergravity with (bosonic) Lagrangian \cite{CJS}
\beqn \label{Leleven}
e^{-1} {\cal L}^{(11)} & = &
\frac{1}{2} R^{(11)} + \frac{1}{4 \cdot 4!}
F_{M_1 \ldots M_4} F^{M_1 \ldots M_4} \non
& & +\,  \frac{1}{12 \cdot 3! (4!)^2}
A_{M_1 M_2 M_3} F_{M_4 \ldots M_7} F_{M_8 \ldots M_{11}} \epsilon^{M_1 \ldots M_{11}}\ ,
\eeqn
where we have adapted the signs to our conventions.
The (bosonic) spectrum only consists of the metric and
a 3-form gauge field.
The Lagrangian (\ref{Leleven}) is the leading order
contribution in a derivative expansion.
One of the next-to-leading-order terms
in this expansion reads \cite{VW,DLM}\footnote{We use the subscript
${\cal L}_1^{(11)}$ here, because there will be further higher derivative terms
relevant in the next section.} 
\be \label{dl1}
e^{-1} {\cal L}_1^{(11)} = -\frac{T_2}{3! 8!}\,
A_{M_1 \ldots M_3} X_{M_4 \ldots M_{11}} \epsilon^{M_1 \ldots M_{11}}
\ee
with
\be
X_8 = \frac{1}{(2 \pi)^4} \left( -\frac{1}{768}
(\mbox{tr} R^2)^2 + \frac{1}{192} \mbox{tr} R^4
\right)  \label{X8}
\ee
and $T_2 \equiv (2 \pi)^{2/3} (2 \kappa_{11}^2)^{-1/3}$ is the
membrane tension. The term (\ref{dl1}) leads to an important
constraint for Calabi-Yau fourfold reductions of
eleven-dimensional supergravity. Due to the relation
\be
\int_{Y_4} X_8\ =\ -\frac{\chi}{24}\ ,
\ee
the term \Ref{dl1} potentially induces a tadpole-term for the 
three-form $A_3$ in compactifications
on Calabi-Yau fourfolds with non-vanishing
Euler number $\chi$, which would render the
resulting vacuum inconsistent \cite{SVW}.
However, the coefficient of the tadpole term
gets further contributions from space-time filling
membranes, non-trivial $F_4$-flux
\cite{BB,SVW,DM} or M5-branes wrapped around
three-cycles in the Calabi-Yau \cite{PM}.
As we do not consider any space-time filling
membranes or wrapped M5-branes here,
the relation
\be \label{tadpol}
\frac{1}{4 \kappa^2_{11}} \int_{Y_4} F_4 \wedge F_4
=\ \frac{T_2}{24}\ \chi
\ee
has to hold for consistency. Postponing the discussion of
non-trivial flux to the next section, we have to consider
fourfolds with $\chi = 0$ for the moment.

Let us now discuss the three-dimensional bosonic spectrum.
Compactifying on a Calabi-Yau fourfold,
from the metric one gets a graviton, $h^{1,1}(Y_4)$
K\"ahler moduli and $h^{3,1}(Y_4)$ complex structure moduli $Z^\alpha$.
As we are interested in discussing the large volume expansion later on 
it is convenient to use rescaled K\"ahler moduli making the  
dependence on the constant background volume explicit. Thus  
the volume of the Calabi-Yau is given by
\be \label{volum}
{\rm Vol}(Y_4) = \vol_0 \tilde \vol(\tilde M_A)\ ,
\ee
where $\vol_0$ is the background volume and $\tilde \vol$ depends
on the rescaled K\"ahler moduli $\tilde M_A$. Furthermore, one expands the 3-form
\be
A_{\mu i\bj} = \sum_{A=1}^{h^{1,1}}
\vol_0^{1/4} A_{A\, \mu}\, e^A_{i \bj} \quad ,
\quad
A_{ij\bar{k}} = \sum_{I=1}^{h^{2,1}}
\vol_0^{3/8} N_{I}\, \Psi^I_{ij\bar{k}} \ ,
\label{ared}
\ee
where we have introduced bases $e^A$ for $H^{1,1}(Y_4)$ and
$\Psi^I$ for $H^{2,1}(Y_4)$. Thus the spectrum comprises 
$h^{1,1}(Y_4)$ vector multiplets and
$h^{2,1}(Y_4) + h^{3,1}(Y_4)$ chiral multiplets. The explicit factors of $\vol_0$ 
have been introduced in (\ref{ared}) because only for fields so
defined,
the kinetic terms stemming from the reduction of the eleven-dimensional
supergravity action are of the same order in the background volume for
all the scalars and vectors. 

In order to derive the low energy effective
action one has to take into account that the Hodge
decomposition $H^3(Y_4) = H^{1,2}(Y_4) + H^{2,1}(Y_4)$
depends on the complex structure.
The basis $\Psi^I$ of $(2,1)$-forms can locally
be chosen to
depend holomorphically on the complex structure
or, in other words,
\be
\bar\partial_{\bar Z^{\bar \alpha}}\Psi^I = 0\ ,\qquad
\partial_{Z^\alpha}\Psi^I \neq 0\ .
\ee
The derivative
$\partial_{Z^\alpha}\Psi^I$ can be expanded into
$(1,2)$- and $(2,1)$-forms with
complex-structure dependent
coefficient functions $\sigma$ and $\tau$
\be 
\partial_{Z^\alpha}  \Psi^I =
\sigma_{\alpha K}^I(Z, \bar{Z}) \,
\Psi^K +
\tau_{\alpha \bar{L}}^I(Z, \bar{Z}) \,
\bar{\Psi}^{\bar{L}}\ .
\label{var}
\ee

Using this when inserting (\ref{ared}) and the usual product 
ansatz for the metric into
(\ref{Leleven}), the three-dimensional effective Lagrangian takes the
form\footnote{Here we performed a Weyl-rescaling $g_{\mu \nu} \rightarrow \tilde \vol^2 g_{\mu \nu}$ 
and a further rescaling of the K\"ahler moduli similar to 
the one in \cite{HL}, i.e.\ $M_A = \tilde M_A \tilde \vol^{-1}$. \label{resc}}
\beqn\label{M3d21}
e^{-1} {\cal L}_0 = \hspace{-.5cm} &&
\frac{1}{2} R
- G_{\alpha \bar{\beta}} \partial_\mu Z^\alpha
\partial^\mu \bar{Z}^{\bar{\beta}} - \frac{1}{2} G^{AB}
\partial_\mu  M_A \partial^\mu M_B  \nonumber \\
&& +\, \frac{1}{4} G^{AB}
F_A^{\mu \nu} F_{B\, \mu \nu} - G^{I \bar{J}}
D_\mu N_I D^\mu \bar{N}_{\bar{J}} \non
&& -\, \frac{1}{8}
 \epsilon^{\mu \nu \rho}
d^{A I \bar{J}} F_{A\, \mu \nu} \left[ N_I
D_{\rho} \bar{N}_{\bar{J}} - D_{\rho} N_I \bar N_{\bar{J}} \right]\ .
\eeqn
The corresponding action is given by $\kappa_3^{-2} \int d^3 x \, {\cal L}_0$ 
and the three-dimensional gra\-vi\-ta\-tional constant is related to the 
eleven-dimensional one by $\kappa_3^{-2} = \kappa_{11}^{-2} (\vol_0 \kappa_{11}^{16/9})$.
Let us explain the notation in (\ref{M3d21}).\footnote{Note that the notation slightly differs
from \cite{HL2} in order to be consistent with the notation of section 2.} 
The scalar sigma-model is described
by the three metrics $G_{\alpha \bar{\beta}}, G^{AB}$ and $G^{I \bar{J}}$.
The Zamolodchikov metric $G_{\alpha \bar{\beta}}$ on the
complex structure moduli space is K\"ahler with
K\"ahler potential
\be \label{kcs}
K_{\rm CS} = -\ln \left( \int_{Y_4} \Omega \wedge \bar \Omega \right)\ ,
\ee
where $\Omega$ is the (4,0)-form of $Y_4$.
Similarly, the metric on the K\"ahler moduli space is
given by the second derivative of a real function
\be \label{kk}
G^{AB} = -\frac12 \partial_{M_A} \partial_{M_B} K_{\rm K}
= -\frac12 \partial_{M_A} \partial_{M_B} \ln \vol\ ,
\ee
where $\vol = \frac{1}{4!} \int_{Y_4} J^4 = \frac{1}{4!} 
d^{ABCD} M_A M_B M_C M_D$ is a rescaled 
volume of $Y_4$, $J=M_A e^A$ being a rescaled K\"ahler form and 
\be \label{int}
d^{ABCD} = \int_{Y_4} e^A \wedge e^B \wedge e^C \wedge e^D
\ee
the quadruple intersection numbers. Also
\be\label{21metric}
G^{I \bar{J}} = - \frac{i}{2} d^{A I \bar{J}}
M_A\ ,
\ee
where we have used the
definition
\be\label{inter}
d^{A I \bar{J}} \equiv \int_{Y_4} e^A \wedge
\Psi^I \wedge \bar{\Psi}^{\bar{J}}\ .
\ee
Furthermore, we introduced the abbreviations
\be
D_\mu N_I = \pu N_I + N_K \sigma^K_{\alpha I}
\pu Z^\alpha + \bar{N}_{\bar{L}}
\bar{\tau}^{\bar{L}}_{\bar{\beta} I}
\pu \bar{Z}^{\bar{\beta}}, \qquad
D_\mu \bar{N}_{\bar{J}} = \overline{D_\mu N_J}\ .
\label{Dmu}
\ee
Let us next make contact with the formulas of the previous chapter.
To do so we have to express (\ref{M3d21}) in the adequate variables,
i.e.\ the chiral fields should be those appearing also in the dualized
version without vector multiplets, and the scalars from the vector
multiplets should obey (\ref{ma}). The correct K\"ahler coordinates
in the dualized theory were determined in \cite{HL2}. In the case
without flux, the dualization can be performed explicitly by adding a
Lagrange multiplier term
\be
e^{-1} {\cal L}_{\rm L.M.} = F^A_\mu \po P_A \quad , \qquad F^{A \rho} = \frac{1}{2}
\epsilon^{\rho \mu \nu} F^A_{\mu \nu}
\label{lagrangemulti}
\ee
and eliminating the fields $F^A_\mu$ via their
equations of motion. The K\"ahler structure of the
resulting sigma-model becomes manifest when expressed
in the coordinates $\phi^i = (Z^\alpha, \hat N^I, \phi^A)$,
where
\beqn
&& \hspace{-1cm} \nN^I = \Gi^{I \bar J}
(Z^\alpha, \bar{Z}^{\bar{\alpha}})\, \bar{N}_{\bar{J}}\ ,\label{Nhat} \\
&& \hspace{-1cm}\phi^A =
i P^{A} - 2 \vol^A \vol^{-1}
+\frac{i}{4} d^{A M \bar{L}} \Gi_{\bar J M}
\Gi_{\bar{L} I}
\nN^{I} \bar{\nN} \, \! ^{\bar{J}}
- \omega^A_{I K} \nN^I \nN^K\ .\label{TA}
\eeqn
Here we have defined a variant of (\ref{21metric})
which is independent of the K\"ahler moduli,
\be\label{hatgij}
\Gi^{I \bar{J}} = - \frac{i}{2} d^{A I \bar{J}} c_A \ ,
\ee
where all $c_A$ can be chosen to be equal to $1$, and
\be
\vol^{A} \equiv \frac{1}{4!}
\int_{Y_4} \oneone^A
\wedge J \wedge J \wedge J\ . \label{vola}
\ee
Furthermore, the $\omega^A_{I K}$ are functions of $Z^\alpha$ and
$\bar{Z}^{\bar{\alpha}}$ which have to obey
\be
\bar\partial_{\bar{Z}^{\bar{\alpha}}} \omega^A_{I K} =
-\frac{i}{4} \Gi_{\bar L I} \Gi_{\bar J K}
d^{A M \bar{L}}
\bar{{\tau}}^{\bar{J}}_{\bar{\alpha} M}\ ,
\label{eqomega}
\ee
but are otherwise unconstrained. In terms of these coordinates the
K\"ahler potential is given by
\be
K = K_{\rm CS} + K_{\rm K}\ ,
\label{m3Dk}
\ee
where $K_{\rm CS}$ and $K_{\rm K}$ are given in (\ref{kcs}) and (\ref{kk})
and have to be expressed in terms of the K\"ahler coordinates $\phi^i$. 
Doing so one realizes that $K$ indeed does not depend on the $\Im(\phi^A)$.  

Now one can show that
\be \label{dtak}
\dd_{\phi^A} K = M_A\ ,
\ee
which indeed coincides with the general formula
(\ref{ma}).\footnote{Notice that $K$ in (\ref{m3Dk})
only depends on the real part $\varphi^A = \Re \phi^A$.
Thus $\dd_{\phi^A} K = \ft12 \dd_{\varphi^A} K$.}
Hence the appropriate variables for making
contact with the last chapter are $\phi^a = (Z^\alpha, \nN^I)$
and  $(M_A, A_{A\, \mu})$.
In the new coordinates the 'covariant' derivatives (\ref{Dmu}) become
\beqn
&& \hspace{-1cm} D_\mu \bar{N}_{\bar{J}} =
\Gi_{\bar{J} I} \pu \nN^I
+ \pu Z^\alpha \left(
\Gi_{\bar{M} K}
\bar{\nN} \, \! ^{\bar{M}}
\tau^K_{\alpha \bar{J}}
- \Gi_{\bar{J} I} \nN^K \sigma^I_{\alpha K} \right)
\ , \non
&& \hspace{-1cm} D_\mu N_I =
\Gi_{\bar{J} I} \pu \bar{\nN} \, \! ^{\bar{J}} +
\pu \bar{Z}^{\bar{\beta}} \left(
\Gi_{\bar{M} K} \nN^K
\bar{\tau}^{\bar{M}}_{\bar{\beta} I}
- \Gi_{\bar{J} I}
\bar{\nN} \, \! ^{\bar{M}} \bar{\sigma}^{\bar{J}}_{\bar{\beta} \bar{M}}
\right)\ ,
\label{Dmutilde}
\eeqn
as can be seen from
\be
\partial_{Z^\alpha} \Gi^{I \bar{J}}
= \sigma^I_{\alpha K} \Gi^{K \bar{J}}\ ,\qquad
\partial_{Z^\alpha} d^{A I \bar{J}}
= \sigma^I_{\alpha K} d^{A K \bar{J}}\ .
\label{dzgij}
\ee
Plugging them into (\ref{M3d21}) one can read off
$G_{a \bar b}$ (which will be given below, in (\ref{GSL})). Moreover, from (\ref{ktilde}) and (\ref{TA})
we see that the kinetic potential $\tilde K$ is given
by\footnote{This corresponds to (\ref{ktilde}) up to an irrelevant constant that
can be absorbed by a K\"ahler transformation of $K$. Moreover, it is very reminiscent
to the K\"ahler potential of the effective two-dimensional theory that one gets 
from a Calabi-Yau fourfold compactification of type IIA \cite{HLM}.}
\beqn \label{kinpot}
&& \hspace{-1.3cm} \tk = K - M_A \left(
\frac{i}{2} d^{AM\bar L} \Gi_{\bar{J} M}
\Gi_{\bar{L} I} \nN^I \bar{\nN} \, ^{\bar J}
-\omega^A_{I K} \nN^I \nN^K - \bar{\omega}^A_{\bar J \bar L}
\bar{\nN} \, ^{\bar J} \bar{\nN} \, ^{\bar L} \right) .
\eeqn
Indeed, one can show
\be \label{gabgsl}
G^{AB} = -\ft12 \partial_{M_A} \partial_{M_B} \tk\ , \quad
G_{a \bar b} = \partial_{\phi^a}
\partial_{\bar \phi^{\bar b}} \tk
\ee
and
\be \label{epsdf}
d^{A I \bar{J}} \left[ N_I
D_{\rho} \bar{N}_{\bar{J}} - D_{\rho} N_I \bar N_{\bar{J}} \right] =
-4 \Im \left[ (\partial_{M_A} \partial_{\phi^a} \tk)
\partial_\rho \phi^a \right] + \partial_\rho (\ldots)\ ,
\ee
where, again, first one has to express the left hand side
through the actual K\"ahler coordinates (\ref{Nhat}).
Using these formulas in (\ref{M3d21}) shows that the low
energy effective Lagrangian ${\cal L}_0$ is indeed of the form
(\ref{Lnew}), taking into account (\ref{dtildek}).

Let us finally mention that it is straightforward but rather tedious to
verify the relations (\ref{Gg}) and (\ref{fdualize}), in the ungauged case.


\subsection{Inclusion of 4-form flux}

In this section we consider the effects of turning
on a background flux for the 4-form $F_4$. The case 
$h^{2,1}(Y_4) = 0$ was already discussed in \cite{HL} and 
here we show that the same result (and calculation) still 
holds in the general case including the (2,1)-moduli $\hat N^I$. 

Let us remind the reader that in order to calculate the potential 
in a Calabi-Yau fourfold compactification with 4-form fluxes one has to 
take into account also one of the higher-derivative terms of the eleven-dimensional 
theory. Only some of them are known explicitly at the moment of writing.\footnote{See 
\cite{PVW} for an overview and for references. We will use the notation 
of \cite{T} in the following.} There is a term

\be \label{dl2} 
e^{-1} {\cal L}^{(11)}_2 = b_1 T_2 \left(J_0 - \frac{1}{2} E_8\right)\ , 
\ee 
where $b_1^{-1}\equiv (2 \pi)^4 3^2
2^{13}$ and 
\beqn \label{e8}
E_8 & = & \frac{1}{3!}
\epsilon^{ABCM_1 N_1 \ldots M_4 N_4}
\epsilon_{ABCM_1' N_1' \ldots M_4' N_4'}
R^{M_1' N_1'}\! _{M_1 N_1} 
\ldots  R^{M_4' N_4'}\! _{M_4 N_4}\ , \\
J_0 & = & t^{M_1 N_1 \ldots M_4 N_4} 
t_{M_1' N_1' \ldots M_4' N_4'} R^{M_1' N_1'}\! _{M_1 N_1}
\ldots  R^{M_4' N_4'}\! _{M_4 N_4} + \frac{1}{4} E_8\ 
.\nonumber
\eeqn
The tensor $t$ is defined by
$t^{M_1 \ldots M_8} A_{M_1 M_2} \ldots A_{M_7 M_8}  
  = 24 {\rm tr} A^4 - 6 ({\rm tr} A^2)^2$
for antisymmetric tensors $A$, and $E_8$ is an eleven-dimensional generalization of the
eight-dimensional Euler density.
More generally one can define \cite{T}
\beqn \label{euler3}
E_n (M_D) & = & \frac{1}{(D-n)!}\, \epsilon_{N_1 \ldots N_{D-n} N_{D-n+1}
\ldots N_D}
\epsilon^{N_1 \ldots N_{D-n} N'_{D-n+1} \ldots
N'_D} \non
& & R^{N_{D-n+1} N_{D-n+2}}\, \! _{N'_{D-n+1} N'_{D-n+2}}
\ldots R^{N_{D-1} N_D}\, \! _{N'_{D-1} N'_D}\ ,
\eeqn
where $D$ denotes the real dimension of the manifold.
Then $E_8(Y_4)$ is proportional to the
eight-dimensional Euler density, i.e.\
\be \label{e8int}
12b_1 \int_{Y_4} d^8 y \sqrt{g}\, E_8(Y_4)  = \chi\ .
\ee
Thus in a Kaluza-Klein reduction 
on a Calabi-Yau fourfold the $E_8$-term in (\ref{dl2}) contributes to the potential of the
three-dimensional effective theory, as has first been noted in
\cite{AFMN}. On the other hand, the integral of $J_0$ over the Calabi-Yau vanishes 
to lowest order in $\kappa_{11}$, i.e. for the Ricci-flat metric \cite{GW}. 

In \cite{HL} it was shown that the contribution of the Euler-term 
combines with the one from the kinetic term of the 4-form to give the 
potential of the effective theory to lowest order in the large volume 
limit. Other possible contributions to the potential might come e.g.\ from (a yet unknown) 
$F_4^2 R^3$-term. However, this would be suppressed in the large volume limit 
as compared to the contributions of $|F_4|^2$ and $E_8$. We will come back to a 
discussion of the limit we are considering at the end of this section. 

Repeating the calculation of \cite{HL} in the case with 
$h^{(2,1)} \neq 0$, we see that a background flux 
$F_4$ leads to the appearance of a potential and a
Chern-Simons term that was first
noted in \cite{PM}. This was to be expected also 
from the general discussion in chapter 2,
cf. (\ref{Lnew}). Thus (\ref{M3d21}) gets corrected by
\be\label{l1}
e^{-1} {\cal L}_g = e^{-1} {\cal L}_0 + \ft12 g
\epsilon^{\mu \nu \rho}\, \Theta^{AB}
 A_{A\, \mu} F_{B\, \nu \rho}\ +\, g^2 V  \ ,
\ee
where $g=\vol_0^{-1/2}$. 
The potential is given by
\be \label{v}
V = e^K G^{\alpha \bar \beta} D_\alpha W
D_{\bar \beta} \bar W +
8 G_{AB} \partial^A T
\partial^B T - 16 T^2\ ,
\ee
where the K\"ahler potential of the gauged theory is the same as the
one of the ungauged theory given in (\ref{m3Dk}). Furthermore,
$D_\alpha = \partial_\alpha + (\partial_\alpha K)$ and the superpotential
$W$ resp.\ the function $T$ are defined as
\be \label{W}
W= \int_{Y_4} \Omega \wedge F_4 \quad , \qquad
T = \frac{1}{16} \int_{Y_4} J \wedge J \wedge  F_4\ .
\ee
Finally, the coefficients of the Chern-Simons term are given by
\be \label{cs}
\Theta^{AB} = 2 \partial^A\partial^B T
= \frac{1}{4} \int_{Y_4} e^A \wedge e^B \wedge F_4\ .
\ee

As we already said the potential is exactly the same as the one calculated 
in \cite{HL} without the $(2,1)$-moduli, as it 
only gets contributions from the kinetic term of the 3-form and
the higher derivative term (\ref{e8}). Moreover, the form of the potential 
(\ref{v}) is perfectly consistent with the general formula (\ref{Vnew}).
Obviously the first line of (\ref{Vnew}) corresponds the last two terms in (\ref{v}). 
Furthermore, the last term of (\ref{Vnew}) vanishes, leading to a
no-scale potential. Finally, to verify that the first term in the 
second line of (\ref{Vnew}) corresponds to the first term in (\ref{v}), we 
notice that $D_I W = 0$ and that the metric $G_{a \bar b}$ has the form
\be \label{GSL}
G_{a \bar b} = \left( \begin{array}{cc}
                   G_{\alpha \bar \beta} + {\cal M}_{K \bar L}
                   {\cal B}^K_\alpha \bar {\cal B}^{\bar L}_{\bar \beta}
                   & {\cal M}_{K \bar J} {\cal B}^K_\alpha \\
                   {\cal M}_{I \bar L} \bar {\cal B}^{\bar L}_{\bar \beta}
                   & {\cal M}_{I \bar J}
                   \end{array} \right)\ ,
\ee
where we have defined
\be \label{Mij}
{\cal M}_{I \bar J} = G^{K \bar L} \Gi_{\bar{L} I} \Gi_{\bar{J} K}\ , \qquad
{\cal B}^I_\alpha = \Gi_{\bar{M} K}
\bar{\nN} \, \! ^{\bar{M}}
\tau^K_{\alpha \bar{L}} \Gi^{I \bar{L}}
- \sigma^I_{\alpha K} \nN^K\ .
\ee
The inverse metric is thus given by
\be \label{invers}
G^{\bar b a} = \left( \begin{array}{cc}
                   G^{\bar \beta \alpha}
                   & - G^{\bar \beta \gamma} {\cal B}^I_\gamma \\
                   - G^{\bar \delta \alpha} \bar {\cal B}^{\bar J}_{\bar \delta}
                   & {\cal M}^{\bar J I} + G^{\bar \delta \gamma}
                     {\cal B}^{\bar J}_{\bar \delta} {\cal B}^I_\gamma
                   \end{array} \right)\ .
\ee
As the $(\bar \beta, \alpha)$-subsector of $G^{\bar b a}$ is just given by $G^{\bar \beta \alpha}$, 
as in the case without the $(2,1)$-moduli $\nN^I$, we see that indeed (\ref{v}) is consistent with 
(\ref{Vnew}).

Let us now comment on the kind of expansion we are making, i.e.\ 
the range of validity of (\ref{l1}). First of all we are considering 
the large volume limit, i.e.\ $\vol_0 \gg 1$. However, 
we also have to restrict the choice 
of the fourfold, in order to end up with (\ref{l1}) as the effective theory. 
For the Chern-Simons term and the potential 
to remain at leading order 
in the large volume limit, we have to consider Calabi-Yau fourfolds 
with large Euler number $\chi$. From (\ref{tadpol}) it is clear that 
the flux is of the order
of the square root of the Euler number, $F_4 \sim \chi^{1/2}$. Thus for
$\chi \sim \vol_0$ 
the Chern-Simons term and the potential are of the same order as the terms in 
${\cal L}_0$.
A second condition on the Calabi-Yau comes from the fact that we want to neglect 
all other contributions from higher derivative terms
that are not contained in (\ref{l1}). As was 
observed in \cite{HL}, the term (\ref{dl2}) leads to a correction to
the three-dimensional 
Einstein-Hilbert term, due to the fact that 
\be \label{eul}
E_8 (M_3 \times Y_4) = E_8 (Y_4) + 4 E_2 (M_3) E_6 (Y_4)\ ,
\ee
where $E_2 (M_3) = 2R$. More precisely, before the Weyl-rescaling 
the three-dimensional Einstein-Hilbert term gets corrected to 
\be \label{ehcorr}
{\cal S}_{\rm EH} = \frac{1}{2 \kappa_3^2} \int d^3x\, e \Big(\tilde \vol + a \vol_0^{-3/4} \int_{Y_4} 
d^8z \sqrt{\tilde g} E_6 (Y_4)\Big) R\ ,
\ee
for some constant $a$. Here we have used the K\"ahler coordinates $\tilde M_A$ of 
(\ref{volum}) again, i.e.\ $\tilde{J}=\tilde{M}_A e^A$.
It can be shown that 
\be \label{c3}
\int_{Y_4} d^8z \sqrt{\tilde g} E_6 (Y_4) \sim \int_{Y_4} \tilde J \wedge c_3\ , 
\ee
where $c_3$ is the third Chern class. Thus the correction term in (\ref{ehcorr})
is the analogue of the 
correction to the Einstein-Hilbert term
proportional to the Euler number found 
for Calabi-Yau threefold compactifications
in \cite{AFMN}. There it was shown that  
the kinetic terms for the moduli get a similar correction. 
Also in our case we expect 
corrections to the kinetic terms $\sim \int_{Y_4} d^8z 
\sqrt{\tilde g} E_6 (Y_4)$, at least 
for the geometrical moduli.\footnote{For the $(2,1)$-moduli 
this would require an
$F_4^2 R^3$-term.} In order for these corrections to be negligible 
in the large volume limit, 
we have to demand that the invariants (\ref{c3}) are small 
for the Calabi-Yau fourfold, which in 
addition should have a large Euler number. 
In \cite{PM,KLRY} the invariants (\ref{c3}) are given for some 
examples, showing that this requirement is a 
rather strong constraint. We have, however, found examples 
in which they are 
considerably smaller than the Euler number.

Two further comments are in order here. First, it was found in
\cite{BB} that a non-vanishing 4-form background in general
necessitates generalizing the metric ansatz to a warped product. 
As argued in \cite{GSS}, the warp factor becomes trivial in the 
large volume limit. Here we are not considering any extended objects 
like space-time filling membranes or wrapped fivebranes, so 
the large volume limit is not obscured by the warp factor 
blowing up close to such sources, hence
we have been working with  the warp factor set to one. 
However, we also explicitly checked that the contributions of the 
warp-factor coming from the Einstein-Hilbert term and from the 
kinetic term of the 4-form in the compactification cancel against each 
other.\footnote{In a compactification
of type IIB with fluxes the effects of the warp factor are discussed in
\cite{dWG}.}

Second, in the presence of a 4-form flux also the expansion
of the higher dimensional fields into harmonic forms, cf.\
(\ref{ared}), requires justification. The motivation to expand
into harmonic forms in the case without flux is that
they lead to zero-modes of the relevant mass operator. However,
this is no longer true in the presence of fluxes. This is another
way of saying that some of the former moduli become massive.
Similarly to the case of the type IIB compactification
on a $T^6$-orientifold with fluxes discussed in \cite{KST}, 
it can be argued here that the ratio of 
the masses due to the flux and
those of the Kaluza-Klein fields is
$(m_{\rm flux} : m_{\rm KK}) \sim (R^{-4} : R^{-1})$,
with $R$ the (dimensionless) 'average' radius of the
Calabi-Yau.\footnote{See also the discussion in \cite{KM}.}
Thus in the large radius limit the flux-induced masses are small
compared to the Kaluza-Klein masses, and it is justified to keep only
the modes coming from an expansion into the harmonic forms in the
low energy effective action.


\subsection{Supersymmetry breaking}

The conditions for unbroken $N=2$ supersymmetry in a Calabi-Yau
fourfold compactification with fluxes were analyzed in
\cite{BB}. There it was found that the flux has to be
of type $(2,2)$ and primitive, i.e. $F_4 \wedge J = 0$.
Taking into account (\ref{W}) this is perfectly consistent
with our general discussion (\ref{n2}).

The case of $N=1$ has not been discussed in the literature so far.
The potential (\ref{v}) is positive definite and can be
written as \cite{BBHL}
\be \label{v2}
V=e^K G^{\Ga {\bar \Gb}}D_{\Ga}WD_{\bar \Gb}{\bar W}
+8 G_{AB}D^A T D^B T\ ,
\ee
with the 'covariant' derivative
$D^A T = \partial^A T - {1 \over 2} (\partial^A K) T$, cf. (\ref{posdef}).
Therefore the condition for unbroken $N=1$ can be read off from (\ref{n1})
to be
\be \label{n1cy}
2 T = \pm e^{K/2}\, |W| \;.
\ee
If this is fulfilled for non-vanishing $T$ and $W$, supersymmetry is
spontaneously broken to $N=1$. To be more precise, it was noted in \cite{GVW,BB2}
that fluxes of type $F_4 \sim J \wedge J$ or $F_4 \sim \Omega + {\rm c.c.}$
lead to a breaking of $N=2$ by generating a non-vanishing $T$ or
$W$, respectively. However, the potential still vanishes. This can be understood by
rewriting the potential (\ref{v}) in yet another way \cite{HL,BBHL}
\be \label{v3}
V = - {\cal V} \Big( \int_{Y_4} F_{3,1} \wedge F_{1,3} + \ft12
\int_{Y_4} J \wedge F_{1,1}^{(0)} \wedge J \wedge F_{1,1}^{(0)} \Big)\ ,
\ee
where we have used the Lefschetz decomposition of a general $(2,2)$-form
flux
\be \label{f22}
F_{2,2} = F^{(0)}_{2,2} + J \wedge F^{(0)}_{1,1} + J^2 \wedge F^{(0)}_{0,0}
\ee
with primitive $(p,p)$-forms $F^{(0)}_{p,p}$. Thus we see that neither
a flux $F_4 \sim J \wedge J$ nor $F_4 \sim \Omega + {\rm c.c.}$ enters
in the potential. Given a (locally constant) flux $F_4$, its expansion
into $(p,4-p)$-forms depends on the complex structure and K\"ahler moduli.
Minimizing the potential (\ref{v3}) restricts the moduli space to the
subspace, where $F_4$ can be expanded as
\be \label{fexp}
\frac{F_4}{2 \pi} = F^{(0)}_{2,2} + F J^2 + F_{4,0} + F_{0,4}\ .
\ee
By writing $F_{4,0} = \tilde F\, \Omega$ the condition (\ref{n1cy})
becomes
\be \label{fft}
\frac{3F}{|\tilde F|} = \pm \sqrt{\frac{\int \Omega \wedge \bar \Omega}{\vol}}\ .
\ee
Note that (\ref{n1cy}) is scale-invariant. Thus if there is one
point in moduli space at which supersymmetry is broken to $N=1$,
then this is true irrespective of the volume $\vol$. This is in
accordance with (\ref{fft}) despite the explicit appearance of the
volume on the right hand side, because the coefficient $F$ has to
scale like $\vol^{-1/2}$ in order to leave the flux (\ref{fexp})
constant.

\subsection{Example}

We now give an example which makes it plausible that it
is indeed possible to fulfill the condition (\ref{fft}).
For simplicity we take the following ansatz for the 4-form flux
\be \label{fans}
\frac{F_4}{2 \pi} = F J^2 + F_{4,0} + F_{0,4}\ ,
\ee
thus discarding any contribution from primitive (2,2)-forms.
The Calabi-Yau fourfold is taken from \cite{KreSka}.\footnote{More precisely,
it is in the first list of reflexive 5d polyhedra.} 
It is a hypersurface $X$ in a weighted projective space, whose weights can 
be found in the (last row of the) table of appendix \ref{table}.
The non-trivial Hodge numbers are given by
\be \label{hodgex2}
h^{(1,1)} = 7 \quad , \quad h^{(1,2)} = 6
\quad , \quad h^{(1,3)} = 1335 \quad , \quad h^{(2,2)} = 5600\ .
\ee
Thus $X$ has Euler number
\be \label{eulerx}
\chi = 8064\ .
\ee
The fact that the Euler number is divisible by $24$ implies that
$H^4(X,\mathbb{Z})$ is an even lattice.\footnote{For a good collection of
general topological properties of Calabi-Yau fourfolds see chapter 2 of
\cite{KLRY}.} Moreover, it was shown in \cite{W} that the flux in general
has to fulfill $F_4/2\pi - {p_1}/{4} \in H^4(X, \mathbb{Z})$, where
$p_1$ is the first Pontryagin class.\footnote{Here we make the
choice $T_2 = 1$, i.e.\ $\kappa_{11}^2 = 2 \pi^2$, in order to make
contact with \cite{W}.\label{t21}} In case $\chi$ is divisible by
$24$ this implies $ F_4/ {2 \pi}\in H^4(X, \mathbb{Z})$.
The way we want to ensure this is by demanding that $J$, $\Re(\Omega)$ and 
$\Im(\Omega)$ are all integral themselves and by using integral 
coefficients $F, \tilde F$ in the expansion (\ref{fans}). 

Thus we assume that we are at a point in moduli space of complex
structure where both $\Re(\Omega)$ and $\Im(\Omega)$ are integral.
As in the case at hand $H^4(X,\mathbb{Z})$ is an even lattice, 
\be \label{omegaint}
\int_X \Re(\Omega) \wedge \Re(\Omega) \ \in 2
\mathbb{Z} \quad , \quad \int_X \Im(\Omega) \wedge \Im(\Omega)
\ \in 2 \mathbb{Z}\ .
\ee
Furthermore, because of $\int \Omega \wedge \Omega = 0$ the two
integrals in (\ref{omegaint}) are actually equal as follows from
\be \label{equal}
\int_X \Omega \wedge \Omega = \left( \int_X
\Re(\Omega) \wedge \Re(\Omega) - \int_X \Im(\Omega) \wedge
\Im(\Omega) \right) + 2 i \int_X \Im(\Omega) \wedge \Re(\Omega)\ .
\ee
This in turn implies that
\be \label{omegax}
\int_X \Omega \wedge \bar \Omega = 2 \int_X
\Re(\Omega) \wedge \Re(\Omega) = 4 r \in 4 \mathbb{Z}\ .
\ee
As this integral is positive one can even be more specific:
$r \in \mathbb{N}$.

Next we choose an integral basis $e^A ( A=1, \ldots, 7)$ for
$H^{(1,1)}(X)$ which is always possible because of $h^{(2,0)}(X) =
0$, i.e.\ $H^2(X,\mathbb{Z}) \otimes \mathbb{C}
= H^{(1,1)}(X)$. Using the Maple packages {\tt puntos} 
\cite{puntos} and {\tt schubert} \cite{schubert} we 
calculate the volume to be
\begin{eqnarray*}
\vol &=&
6\,M_2\,M_3\,M_4\,M_5 + 
{\textstyle \frac {8}{3}} \,M_4\,M_5^{3} + 
{\textstyle \frac {4}{3}} \,M_2\,M_5^{3} + 6\,
M_3\,M_4\,M_5^{2}  
 \\ &&
\mbox{} + 12\,M_3\,
M_5\,M_4^{2} + 8\,M_3\,M_6^{3} + {\textstyle \frac {8}{3}} \,M_5\,M_4
^{3} + 2\,M_1\,M_2\,M_5\,M_6  \\ &&
\mbox{} + 2
\,M_1\,M_2\,M_6^{2} + 6\,M_1\,
M_5^{2}\,M_6 + 4\,M_1\,M_5\,
M_6^{2} + {\textstyle \frac {8}{3}} \,M_1\,M_6
^{3}  \\ &&
\mbox{} + 6\,M_1\,M_7^{3} + 2\,M_2\,{
M_4}\,M_5^{2} + 3\,M_5^{2}\,M_6^{2} + 
{\textstyle \frac {2}{3}} \,M_1^{3}\,M_7 + 3\,
M_2\,M_7^{3}  \\ &&
\mbox{} + {\textstyle \frac {8}{3}} \,M_5\,M_6
^{3} + 12\,M_3\,M_5\,M_6^{2} + 18\,
M_3\,M_5^{2}\,M_6 + 2\,M_6^{4} + 
{\textstyle \frac {4}{3}} \,M_2\,M_6^{3} 
 \\ &&
\mbox{} + 6\,M_2\,M_3\,M_5\,M_6
 + M_2\,M_5\,M_6^{2} + 6\,M_2\,
M_3\,M_6^{2} + 2\,M_1\,M_2\,
M_4\,M_5  \\ &&
\mbox{} + 5\,M_5^{2}\,M_4^{2} + {\textstyle \frac {9}{2}} \,M_7^{4} + 6\,
M_2\,M_3\,M_5^{2} + {\textstyle 
\frac {2}{3}} \,M_5^{4} + M_2\,M_5\,
M_4^{2}  \\ &&
\mbox{} + {\textstyle \frac {2}{3}} \,M_4^{4} + 
{\textstyle \frac {5}{6}} \,M_1^{4} + {\textstyle 
\frac {32}{3}} \,M_1^{3}\,M_3 + 48\,M_1^{
2}\,M_3^{2} + 12\,M_1\,M_2\,M_3^{
2} \\ &&
\mbox{} + 8\,M_3\,M_4^{3} + 2\,M_1\,M_4\,M_5^{2} + 
4\,M_1\,M_5\,M_4^{2} +  M_1^{2}\,M_2\,M_7 \\ &&
\mbox{} + 3\,M_1\,M_2\,M_7^{2} + 3\,M_1^{2}\,{
M_7}^{2} + 8\,M_3\,M_5^{3} + {\textstyle 
\frac {8}{3}} \,M_1\,M_5^{3} + 2\,M_1\,
M_2\,M_5^{2} \\ &&
\mbox{} + 4\,M_1^{2}\,M_2\,M_3 + 
{\textstyle \frac {1}{3}} \,M_1^{3}\,M_2 + 96
\,M_1\,M_3^{3} + 72\,M_3^{4} + 12\,
M_2\,M_3^{3}  \\ &&
\mbox{} + 2\,M_1\,M_2\,
M_4^{2} + {\textstyle \frac {8}{3}} \,M_1\,M_4
^{3} + {\textstyle \frac {4}{3}} \,M_2\,M_4^{3
} + 6\,M_2\,M_3\,M_4^{2}\ . 
\end{eqnarray*}

Now we have to make sure that both the condition (\ref{fft})
and the tadpole condition
(\ref{tadpol}) are fulfilled at the same time. For the choice 
$T_2 =1$  that we use for this
example (see footnote \ref{t21}) (\ref{tadpol}) takes the form
\be \label{tadpol2}
\frac{1}{4 \pi^2} \int_{Y_4} F_4 \wedge F_4 = \frac{\chi}{12}\ .
\ee
Inserting $\frac{F_4}{2 \pi} = F J^2 + (\tilde F \Omega + c.c)$
and using (\ref{fft}) leads to the two conditions
\be \label{twocond}
F^2 = \frac{\chi}{504 \vol} \quad , \quad |\tilde F|^2 = \frac{\chi}{224r} \ .
\ee
Assuming $r=1$ and choosing the K\"ahler moduli
\be \label{km}
M_1 =2\quad , \quad M_2 = 1\quad , \quad M_3 = \ldots = M_7 = 0\ ,
\ee
leading to a volume of $\vol = 16$ and an integral K\"ahler form, 
we see that for the value of the Euler number (\ref{eulerx}) these two conditions 
can be satisfied by
\be \label{FFT}
F = 1 \quad , \quad \tilde F = 6\ .
\ee
Thus we see that at this special point in moduli space, the flux 
\be \label{f4}
\frac{F_4}{2 \pi} = J \wedge J + 12 \Re\, (\Omega) 
\ee
is in $H^4(X, \mathbb{Z})$, fulfills the tadpole condition and breaks supersymmetry
from $N = 2$ to  $N = 1$. 

Strictly speaking the values (\ref{km}) for the 
K\"ahler moduli are outside the range of validity of our effective 
theory (which is at 
large volume), because some of the K\"ahler moduli vanish. As this
means 
that some of the 
two-cycles of the Calabi-Yau shrink to zero size, one generically
expects 
new light 
degrees of freedom from membranes wrapped around these vanishing cycles.
This problem is due to our special Ansatz for $F_4$. To ensure
integrality 
of $\frac{F_4}{2 \pi}$,
we demanded that $J$, $\Re(\Omega)$ and $\Im(\Omega)$ are all integral
themselves.
Relaxing this requirement would certainly allow for 
partial supersymmetry breaking within the range of large volume. 

Let us look at bit closer at this issue. Giving up the demand for an 
integral $J$ and requiring only 
$F J^2 \in H^4(X, \mathbb{Z})$, shows that one is free to rescale all 
K\"ahler moduli 
by a common factor $\sqrt{\lambda}$ (and thus rescaling the volume ${\cal V}$ by 
$\lambda^2$) if one rescales $F$ by $\lambda^{-1}$ at the same time. This 
corresponds to the observation made below (\ref{fft}). However, even 
then some of the K\"ahler moduli 
remain at zero. Thus to avoid 
this problem one should give up also the requirement that $F J^2$ and 
$(\tilde F \Omega + c.c.)$ are separately in $H^4(X, \mathbb{Z})$, a 
situation which is 
hard to analyze explicitly.

\bigskip

\noindent
{\bf Acknowledgments}

We are grateful for useful discussions with 
B.~K\"ors, S.~Hellerman and M.~Schulz.
We would especially like to thank V.~Braun and J.~de Loera 
for essential help with the Maple packages {\tt puntos} and {\tt schubert}.

This work was supported in part by
I.N.F.N., by the EC contract HPRN-CT-2000-00122, by the EC contract
HPRN-CT-2000-00148, by the INTAS contract 99-0-590 and by the MURST-COFIN
contract 2001-025492. M.B.\ was supported by a Marie Curie Fellowship, contract
number HPMF-CT-2001-01311.

\appendix \noindent

\renewcommand{\theequation}{\Alph{section}.\arabic{equation}}
\setcounter{equation}{0}
\setcounter{section}{0}

\section{Search for an example}

To satisfy the $N=1$ condition \Ref{fft} with $r$ in \Ref{omegax}
being unity, we need a Calabi-Yau fourfold with 
an Euler number $\chi$ divisible 
by 24 and 224, and a volume $\vol$ which is
such that $\chi/(504 \vol)$ is a square integer
for some values of the K\"ahler moduli  (cf.\ \Ref{twocond}). 
For example, for $\chi=2016$, this means $\vol$ can be 1 or 4.
If none of these special volume values can be obtained for 
integral K\"ahler moduli $M_A$,
the Calabi-Yau manifold does not satisfy the 
$N=1$ condition \Ref{fft}
given tadpole cancellation and the integrality
assumptions of section 3.

Now, in the tables \cite{KreSka} of Calabi-Yau fourfold 
hypersurfaces in toric varieties, 
the Euler number and a few other properties are explicitly given.
To compute the volume $\vol$ in terms of the K\"ahler moduli,
one also needs intersection numbers of the corresponding
divisors, and for this one needs to desingularize the toric variety. 
This is done by triangulating the corresponding polytope
(see for instance \cite{Greene} for a review).
Triangulating our way 
 through the tables \cite{KreSka} in order of increasing 
number of points in the polytope and increasing
$h^{(1,1)}$, it turns out that three of the candidate toric varieties 
we tried (indicated in table 1) cannot be desingularized;
fractional intersection numbers appear.
In the other five cases, the toric variety can be made smooth,
and in the last candidate in the table,
the condition \Ref{fft} can, in fact, be satisfied
together with tadpole cancellation and integrality conditions ---
for the values in \Ref{km} only.

\[
\begin{array}{|c|c|c|c|c|c|c|c|c|c|c|c|c|c|} \hline
\multicolumn{6}{|c|}{\mbox{weights}} & \mbox{degree}
& \mbox{M } & \mbox{N } & h^{(1,1)}, h^{(1,2)}, h^{(1,3)}
& \chi & \mbox{comment} \\ \hline \hline
1 & 2 & 3 & 3 & 3 & 6 & 18 & 354 & 9 & 3,0,325 &
2016 & \mbox{singular} \\ \hline
1 & 2 & 4 & 7 & 7 & 7 & 28 & 367 & 10 & 4,9,333 &
2016 & \mbox{singular} \\ \hline
1 & 1 & 1 & 1 & 4 & 7 & 15 & 1525 & 10 & 4,0,1332 &
8064 &  \\ \hline
1 & 1 & 2 & 2 & 6 & 12 & 24 & 1547 & 10 & 4,0,1332 & 
8064 & \\ \hline
2 & 2 & 2 & 2 & 7 & 15 & 30 & 1021 & 13 & 4,0,1332 &
8064 &  \\ \hline
1 & 2 & 2 & 2 & 8 & 15 & 30 & 1461 & 14 & 5,0,1331 &
8064 & \\ \hline
1 & 2 & 2 & 6 & 10 & 21 & 42 & 1483 & 12 & 4,0,1332 &
8064 &  \mbox{singular} \\ \hline
1 & 1 & 6 & 8 & 8 & 24 & 48 & 1547 & 13 & 7,6,1335 &
8064 & \mbox{$N=1$}   \\  \hline
\end{array}
\]
Table 1: \label{table} Some Calabi-Yau hypersurfaces
in weighted ${\mathbb P}^{n_1, \ldots n_6}$,
with Euler numbers divisible by 24, 224 and 504, 
taken from the tables of \cite{KreSka}. The 
number of points in the M and N lattices
are given; the N-lattice polytope is the one 
that is triangulated.

\end{document}